# Growth and characterization of the magnetic topological insulator candidate $Mn_2Sb_2Te_5$

Ankush Saxena[1,2], and V.P.S. Awana[1,2*]

[1]*Academy of Scientific & Innovative Research (AcSIR), Ghaziabad-201002*

[2]*CSIR- National Physical Laboratory, New Delhi-110012*

**Abstract:**

We report a new member of topological insulator (TI) family i.e., $Mn_2Sb_2Te_5$, which belongs to $MnSb_2Te_4$ family and is a sister compound of $Mn_2Bi_2Te_5$. An antiferromagnetic layer of $(MnTe)_2$ has been inserted between quintuple layers of $Sb_2Te_3$. The crystal structure and chemical composition of as grown $Mn_2Sb_2Te_5$ crystal is experimentally visualized by single crystal XRD (SCXRD) and field emission scanning electron microscopy (FESEM). The valence states of individual constituents i.e., Mn, Sb and Te are ascertained through X-ray photo electron spectroscopy (XPS). Different vibrational modes of $Mn_2Sb_2Te_5$ are elucidated through Raman spectroscopy. Temperature-dependent resistivity $\rho(T)$ of $Mn_2Sb_2Te_5$ resulted in metallic behaviour of the same with an up-turn at below around 20K. Further, the magneto-transport $\rho(T)$ vs H of the same exhibited negative magneto-resistance (MR) at low temperatures below 20K and small positive at higher temperatures. The low Temperature -ve MR starts decreasing at higher fields. The magnetic moment as a function of temperature at 100Oe and 1kOe showed AFM like down turn cusps at around 20K and 10K. The isothermal magnetization (MH) showed AFM like loops with some embedded FM/PM domains at 5K and purely paramagnetic (PM) like at 100K. The studied $Mn_2Sb_2Te_5$ clearly exhibited the characteristics of a magnetic TI (MTI).

**Keywords**: Magnetic Topological Insulators, Crystal Growth, Raman Spectroscopy, X-ray Photoelectron Spectroscopy, Magneto-resistance, Magnetization.

PACS: 74.70.Dd, 74.62.Fj, 74.25.F-

*Corresponding Author

Dr. V. P. S. Awana:  E-mail: awana@nplindia.org
Ph. +91-11-45609357, Fax-+91-11-45609310
Homepage: awanavps.webs.com

**Introduction:**

The intrinsic combination of anti-ferromagnetism (AFM) within topological insulators (TI) is expected to exhibit various novel phenomenon such as anomalous quantum hall effect (AQHE) [1-3] and are termed as magnetic topological insulator (MTIs). Most of the MTIs are grown through doping or magnetic layer insertion method. It is an uphill task to grow intrinsic MTI and to observe the novel phenomena of AQHE in such materials. Till date only few MTIs are known such as $MnBi_2Te_4$ [4,5], $MnSb_2Te_4$ [6,7], $FeBi_2Te_4$ [8] and $Mn_2Bi_2Te_5$ [9] etc. All these discovered MTIs exhibit the low temperature AFM along with the TI character. Theoretically it is predicted that $MnBi_2Te_4$ have an antiferromagnetic topological insulating state [10-12], which was soon confirmed through experimental studies [13-15]. The previous reports on $MnSb_2Te_4$ suggest that it is a A-type antiferromagnetic material [16], in which a ferromagnetic constituent grows in $MnSb_2Te_4$ below 25K. It is observed through Density functional theory (DFT) calculations that $MnSb_2Te_4$ is expected to be a TRS broken type-II Weyl semimetal.

In the septuple layer (SL) $A^{IV}(Bi/Sb)_2Te_4$ compound the $(A^{IV}Te)_n$ sublattice is built between the QL $(Bi/Sb)_2Te_3$ vdW blocks. The reports on others materials of this series such as $Sn_2Bi_2Te_5$ [17] and $Ge_2Sb_2Te_5$ [18] suggesting that these formed by hexagonal nonuple layer (NL) building blocks. Theoretically it is predicted that in NL $A^{IV}_2(Bi/Sb)_2Te_5$ compounds, the Mn spins couple ferromagnetically in each layer although Mn layers couple antiferromagnetically together in z-plane [19]. Experimentally it is found that the MTIs $Mn_2Bi_2Te_5$ has NiAs-type ABAC stacking within the $(MnTe)_2$ sublattice, in which the FM Mn layers are coupled together antiferromagnetically between NL block [20]. Recently a group of Lin Cao *et.al* reported an experimental report on sister compound $Mn_2Bi_2Te_5$ having a large dynamical axion field [9]. The study of $Mn_2Bi_2Te_5$ is suggesting that it consists rich magnetic topological quantum phenomena i.e., quantum hall effect. The observance of AQHE in series of such MTIs experimentally is yet a challenging task. The reports on $Mn_2Bi_2Te_5$ suggested that it host a huge dynamical axion field [21]. Interestingly, such $A^{IV}(Bi/Sb)_2Te_4$ and $A^{IV}_2(Bi/Sb)_2Te_5$ materials exists a phenomenon namely AQHE, which could only be observed at very low temperature after setting in of the magnetic order [19,23]. It is yet a challenging task for condensed matter physicists or chemists, to discover a near room temperature MTI, i.e., both magnetic order and topological character existing right up to room temperature. It is important to discover new MTIs, having simultaneous existence of nontrivial topological character along with magnetic ordering at close to room temperature to design the advanced spintronics devices and the observance of AQHE at elevated temperatures.

In series of MTIs, here we are reporting a new probable MTI i.e., $Mn_2Sb_2Te_5$. This is first experimental report on single crystalline $Mn_2Sb_2Te_5$. Here we successfully grew the single crystal of $Mn_2Sb_2Te_5$ by the self-flux vacuum encapsulation method. To grow the $Mn_2Sb_2Te_5$, we have inserted two layers of MnTe between the QL of $Sb_2Te_3$ TIs which has same stacking as reported for $Mn_2Bi_2Te_5$. Interestingly, this material has only a single building block having vdW gaps and is realized as intrinsic antiferromagnetic (AFM) material experimentally. Various characterisations such as structural, electrical, magnetic and magneto-transport are performed on as grown single crystalline $Mn_2Sb_2Te_5$ and the results are reported here.

**Experimental:**

$Mn_2Sb_2Te_5$ single crystal is grown by vacuum encapsulation through self-flux method. Premium sigma aldrich powders are taken in stoichiometric content of Mn (99.999%), Sb (99.999%) and Te (99.999%). The powder mixtures are then ground in an MBraun glove box in argon atmosphere with an agate mortar and pestle to obtain a homogeneous mixture. The homogeneous mixture is then pelletized at 100 N/cm$^2$ hydraulic pressure in a circular die. Further, the as obtained pellet is sealed in a quartz bulb at a pressure of $5*10^{-5}$ mbar under vacuum. The $Mn_2Sb_2Te_5$ sample tube was heated to 1000°C at 120°C/hr using a PID controlled oven. The oven is kept at this temperature for up to 48 hours, followed by very slow cooling rate (5°C/hour) to 600°C in over 120 hours for crystallization and maintained at this temperature for 12 hours again for better shining and crystallinity. The sample tube is subsequently quenched in room temperature water from 600$^0$C and finally taken out. A schematic diagram of the heat treatment for $Mn_2Sb_2Te_5$ single crystal growth is shown in Fig. 1. The obtained silvery shinning crystal is shown in the inset of Fig. 1.

Crystalline XRD of small $Mn_2Sb_2Te_5$ flakes was taken with a Rigaku desktop X-ray diffractometer (XRD) equipped with Cu-K$_\alpha$ radiation at a wavelength of 1.5418Å. VESTA-based software was used to design the unit cell for the synthesis of $Mn_2Sb_2Te_5$ single crystals. SEM and EDS analysis were performed with the help of Jeol JSM 7200F FESEM. The Raman spectra of the sample is taken on a Renishaw Reflex Raman microscope equipped with 514 nm and 720 nm lasers. Chemical Analysis of Amorphous $Mn_2Sb_2Te_5$ is executed through Nexsa-thermo fisher scientific XPS by using Al $K_\alpha$ X-ray source.

Transport studies were carried on the Quantum Design Physical Property Measurement System (QD-PPMS) at various temperatures from 2.5 K to 150 K in the presence of an applied external magnetic field of ± 14T. The magnetic properties are carried out on quantum design magnetic properties measurement system (QD-MPMS) in presence of external magnetic field of up to 1.2T and in temperature range 100K down to 2K.

**Results and Discussion**

The SCXRD spectrum of a flake of $Mn_2Sb_2Te_5$ is shown in Fig. 2(a). In SCXRD, all peaks can be identified from the (0 0 L = 1,2,3….) reflections, confirming that the exposure surface of $Mn_2Sb_2Te_5$ is in the ab plane. Although first MTIs $Mn(Bi/Sb)_2Te_4$ has the reflection plane along (0 0 3L) as reported earlier [5] even $Mn_2Sb_2Te_5$ grows along (0 0 L) plane. This shows that the synthesized $Mn_2Sb_2Te_5$ crystal is grew unidirectionally along the c-axis. All the observed peaks are indexed with their respective planes. The unit cell of single crystal is shown in fig.2(b) design through VESTA software by considering the lattice parameter a=b=4.288 Å and c=17.15 Å reported earlier [9] for the sister MTIs $Mn_2Bi_2Te_5$ which also consist a NL block. Due to the magnetic coupling between Mn layers in NL block, the magnetic state with opposite alignment of ferromagnetic layer coupled with Mn spins of nearest NL as shown in fig.2(b). Experimentally it is found that the c parameter is near to 40-42Å for MTIs $A^{IV}(Bi/Sb)_2Te_4$, and 17.15Å for as grown MTIs $Mn_2(Bi/Sb)_2Te_5$. The change is due to insertion of atomic layer of $(MnTe)_2$. The parental SL $A^{IV}(Bi/Sb)_2Te_4$ compound has SL in ABC static sequence with three blocks [23]. In case of $Mn_2Sb_2Te_5$, the Mn layers are coupled together antiferromagnetically between NL block and suggesting an NiAs-type ABC stacking within the $(MnTe)_2$ sublattice. The static sequence can be termed as Te-Sb-Te-Mn-Te-Mn-Te-Sb-Te which is in good agreement with earlier report on $Mn_2Bi_2Te_5$ [20].

FESEM image of single crystal $Mn_2Sb_2Te_5$ is shown in Fig.3(a). Linear growth along the c-axis is visualized at 2 μm resolution. The atomic composition of the elements of single crystal $Mn_2Sb_2Te_5$ is shown in Fig. 3(b). The Quantitative energy dispersive X-ray (EDX) analysis indicates that the atomic percentage for Mn, Sb and Te is 24.54%, 25.24%, and 50.22% respectively, which is close to that of the stoichiometric formula. Both structural (SCXRD) and microstructural (FESEM/EDS) approve the crystallinity (0 0 L), morphology and the composition of studied as grown proposed MTI $Mn_2Sb_2Te_5$ crystal.

The vibrational modes of as grown MTIs Mn$_2$Sb$_2$Te$_5$ are elucidated through Raman spectroscopy. The scattered Raman spectroscopy was carried out at room temperature using a Raman spectrometer in the backscattering configuration excited by He-Ne laser ($\lambda$ = 632.8 nm) and a solid-state green laser ($\lambda$ =532 nm). The primitive unit cell (PUC) of Mn$_2$Sb$_2$Te$_5$ contains nine atoms. Consequently, there are 27 lattice dynamical modes at the center of the Brillouin zone, three of which are acoustic modes and 24 are optical modes. The classification of group theory implies that out of these 24 optical modes, four optical modes are Raman-active modes as $4A^1_{1g}$, $4E^1_g$ symmetrically, and four are infrared-active modes as $4A^1_{1u}$ and $4E^1_{1u}$ symmetrically, and the intricate depictions for the optical phonons can be written as $\Gamma_{optical}$ =4 $A^1_{1g}$+4 $E^1_g$+4 $A^1_{1u}$ +4 $E^1_u$ where $E^1_g$ and $E^2_u$ modes are doubly degenerate [25]. The lowest frequency vibrational mode $E^1_g$ has not been identified on the lattice vibrations of Mn$_2$Sb$_2$Te$_5$ due to dominating $A_{1g}$ mode. The laser power of below 70 μW was focused at the sample surface to avoid the burning of the sample. A power meter is used to measure the intensity of laser beam through a 10x objective.

The fitted experimental data is elucidated in fig.4. Five Raman modes are observed at 61±1 cm$^{-1}$, 86±1 cm$^{-1}$, 116±1 cm$^{-1}$, 117±6 cm$^{-1}$, and 137±1 cm$^{-1}$ respectively through Lorentz deconvolution. All the observed peaks are different from as reported in mother compound Mn(Bi/Sb)$_2$Te$_4$ [24, 25]. Generally, in case of $A^{IV}$(Bi/Sb)$_2$Te$_4$, five peaks are observed in which high frequency vibrational modes is observed between 50 to 120 cm$^{-1}$. Interestingly here five peaks of Mn$_2$Sb$_2$Te$_5$ suggested that it is also a family member of $A^{IV}$(Bi/Sb)$_2$Te$_4$ MTIs. The highest frequency vibrational mode is observed at 116±1 cm$^{-1}$ and termed as $E^2_g$ mode. The shifting in vibrational modes confirms the successfully insertion of (MnTe)$_2$ layers in unit cell of TIs Sb$_2$Te$_3$ [22]. It is also interesting that the middle atomic layer is Mn for MnSb$_2$Te$_4$, which is replaced by Te layer in middle in case of Mn$_2$Sb$_2$Te$_5$. All the observed peaks are frequency dependent and can be termed as low frequency vibrational mode ($A^1_{1g}$, $E^1_g$), middle frequency vibrational mode ($A^2_g$, $E^2_g$), higher frequency vibrational mode ($A^3_g$, $E^3_g$). The similar Raman modes are observed in some isostructural compounds of $A^{IV}$(Bi/Sb)$_2$Te$_4$ family i.e., MnSb$_2$Te$_4$ [25], SnBi$_2$Te$_4$ [39] and SnSb$_2$Te$_4$ [40]. The low-frequency $A^1_{1g}$ and $E^1_g$ modes consist out-of-plane vibrations of Sb and Te atoms in a-b plane [40]. Due to which the middle layer of Mn remains intact. Such modes are very similar to the $A^1_{1g}$ and $E^1_g$ modes exists in Sb$_2$Te$_3$ [22]. In which, Te and Sb atoms vibrate in the same phase. The existence of middle frequency vibrational modes $E^2_g$ at 86.57±0.5cm$^{-1}$, and $A^2_g$ at 117.83±6.1cm$^{-1}$ is due to presence of inserted twice layer of MnTe in Sb$_2$Te$_3$ which suggest a successful insertion of layer in TIs.

The XPS spectra of as grown MTIs Mn$_2$Sb$_2$Te$_5$ is fitted well with the mixing of Gaussian–Lorentzian components. The XPS survey spectrum of MTIs is observed for the binding energy range from 500 to 700 eV, as shown in Figs. 5(a), (b) and (c). The survey spectra of as grown MTIs Mn$_2$Sb$_2$Te$_5$ samples show that Mn, Sb, Te, elements coexist, and no other impurities were present in the samples. Fig.5(a, b, c) indicates the high-resolution scan for Mn (2p), Sb (3d), Te (3d), respectively.

Fig.5(a) shows the high resolution XPS spectrum of the deconvoluted Mn 2p i.e., 2p$_{3/2}$, and 2p$_{1/2}$ with its three sub peaks as shown in fig. The strongest peaks of Mn-2p$_{3/2}$ and Mn-2p$_{1/2}$ peaks at 641.0eV and 653.1 eV are due to the contribution of Mn$^{2+}$ in Mn-Te broken bonds which elucidates the major oxidation state of Mn is 2$^+$. A small separation between both 2p is found to be 12eV although it is just close to the reported by other [26]. The other peaks of Mn$^{3+}$ associated with Mn-2p$_{3/2}$ and Mn-2p$_{1/2}$ identified at 642.5 eV and 654.8 eV, and Mn$^{4+}$ associated with Mn-2p$_{3/2}$ and Mn-2p$_{1/2}$ peaks identified at 645.0 eV and 657.9 eV are known as satellite peaks of Mn valence states. Due to the charge transfer between 3d unfilled subshell and outermost shell of Te, such type satellite peaks exist in photoelectron process. The existence behind such satellite peaks is similar to MnBi$_2$Te$_4$ [27,28].

Fig.5(b) shows the XPS peaks of Sb-3d characterized by a doublet and exists due to spin–orbit splitting (3d$_{3/2}$ and 3d$_{5/2}$). The first peak is found at 539.94±0.04 eV (Sb 3d$_{3/2}$) and other one at 531.61±0.02 eV (Sb 3d$_{5/2}$) respectively. Other two shoulder peaks observed at 538.24±0.10 eV and 529.10±0.01 eV respectively are due to the short exposure to air [29,30]. These oxide peaks are corresponding to Sb-O bonds. Sb$^{3+}$ is identified as the chemical state of antimony. The observed peaks of doublets and air exposer are in well agreement as reported earlier for Sb-3d shell [31].

Fig.5(c) shows the high resolution XPS spectra for Te (3d) which arises as a doublet due to spin orbit coupling of Te 3d$_{3/2}$ and 3d$_{5/2}$ peaks. Furthermore, the Te 3d peak contains four sub-peaks. Two of them represent the MnSb$_2$Te$_4$ phase, centered at 573.21±0.01 eV (Te-3d$_{7/2}$) and 583.62±0.02 eV (Te-3d$_{5/2}$), in Sb-Te and Mn-Te bonding. To elucidates the oxidation exposure the single crystal of Mn$_2$Sb$_2$Te$_5$ is exposed in air for some days. Due to the air exposer the two peaks are observed at 576.45±0.02eV and 587.62±0.01 eV, corresponds to $Te^{2-}$ ($TeO_2$) due to the surface oxide. There is just small shift in peaks with respect to binding energy (B.E) as reported by other for $MnSb_2Te_4$ [32]. The reported B.E. for the TIs $Sb_2Te_3$ (Te-3d) are 572.0 eV and 582.3 eV corresponding to Te $3d_{5/2}$ and Te $3d_{3/2}$ respectively [33-35]. Although the surface oxide peak ($TeO_2$) is identified at 574eV. The shift in B.E. is mainly is due to presence of double magnetic layer of MnTe in TIs $Sb_2Te_3$. Overall, the XPS of exfoliated crystal $Mn_2Sb_2Te_5$ suggested that MnTe layer is successfully inserted between the septuple layer of $MnSb_2Te_4$. All observed XPS peak positions of as grown MTIs $Mn_2Sb_2Te_5$ and Full-width at half maximum (FWHM) of constituent elements is mentioned in table 1.

The temperature dependent longitudinal resistivity in absence of external magnetic field is shown in fig.6. The resistivity starts to decrease as temperature falls off. The longitudinal resistivity in temperature range 20K<T<250K elucidates the highly metallic nature. The longitudinal resistivity is near to 17.8 mΩ–cm at 250K and decreased to 14.3 mΩ–cm at 2K. Below 20K the resistivity shows the semiconductor like behavior although the resistivity takes again an upturn which is due to change in spin-flip scattering of $Mn^{2+}$ moments. Similar effect can be seen in other metallic materials reported by others [36]. Such behavior of resistivity suggests the strong spin disordering which further can be seen in magnetoresistance nature.

The transverse magneto-resistance percentage (MR%) as a function of field of the single crystal $Mn_2Sb_2Te_5$ is shown in fig.7(a) and 7(b). The MR% is calculated by using the resistivity dependent formula

$$MR\% = \frac{\rho(H)-\rho(0)}{\rho(0)} *100$$

Here, ρ(H) is field dependent resistivity, ρ(0) is zero field resistivity. The MR curve elucidates the magnetic state of system. The observation of negative magneto-resistance (NMR) in nonmagnetic materials like as topological semimetals (TSM) is rare [37-48], which is signature that TSM host chiral anomaly [49-51]. Although the observation of NMR in TIs in which chiral anomaly is not defined created a huge confusion in condensed matter physics [52-57]. Here we have observed the NMR at low field ±2T, and the NMR is found to be -4.4% at 2.5K which is in well agreement with its sister MTIs $Mn_2Bi_2Te_5$ at low field [9]. The negative MR of cusp-shape at low fields is analogous to that expected for the weak localization [58,59]. Such type of nature of negative MR is also observed in other TIs

SnBi$_2$Te$_4$ [60] and SnSb$_2$Te$_4$ [61] which appear due to presence of weak anti-localization effect.

Interestingly, the MR, varies with temperatures which is observed negatively below 50 K, and increases from -4.4% to -1% as T is varied from 2.5 K to 50 K. In case of MnSb$_2$Te$_4$, the temperature dependent MR% is just opposite to that of the weak localization (WL) effect, where quantum coherence is responsible for resistance erection, which become weaker at higher temperatures. Therefore, the WL is unlikely responsible for the negative MR in A$^{IV}$(Bi/Sb)$_2$Te$_4$ [62,63]. In comparison with MnSb$_2$Te$_4$, the as grown MTIs Mn$_2$Sb$_2$Te$_5$ implies a positive MR at higher temperature. This huge change in MR of Mn$_2$Sb$_2$Te$_5$, exists due to spin orientations at the surface and interactions between electrons in the presence of external magnetic field. The increase in resistance with field exists due to trajectory of charge carriers which cause classical MR arises which is commonly seen in parental MTIs MnBi$_2$Te$_4$ [63].

The negative MR below T$_N$ is signature of the strongest spin-disordered although the carrier scattering is effectively suppressed in this region. Such type behavior is reported in similar type of materials [64,65]. On the other hand, the -ve MR also persist well beyond the magnetization hysteresis loop in presence of field. This indicates that the negative spin after negative MR cannot be attributed to dynamic walls. This may be related to atomic-level defects such as randomly distributed Mn spins in Sb substitutions. Because of the local disorder, it will not be as good compared to most of the Mn spin in the midplane of the Mn$_2$Sb$_2$Te$_5$ nonuple layer.

The magnetization vs temperature (M-T) plot of as grown Mn$_2$Sb$_2$Te$_5$ is shown in fig.8(a). The M-T measurements have been carried out at 100 Oe and 1kOe under field cooled (FC) and zero field cooled (ZFC) measures. The magnetic moment of Mn$_2$Sb$_2$Te$_5$ in (H∥ab) is found to be zero and remains constant till 50K from 160K. The FC and ZFC plots show a bifurcation below 20 K at 100Oe and below 11K at 1kOe. The bifurcation between ZFC and FC at 100Oe in H∥ab plane suggesting a disordered spin glass state below the freezing temperature T$_f$ ~ 20 K [66,67]. Such type of spin glass behavior is reported earlier for some other materials [68-70] and for its mother compound MnSb$_2$Te$_4$[32]. At 1kOe, both FC and ZFC plots at below 20 K elucidating the ferromagnetic ordering in Mn$_2$Sb$_2$Te$_5$ single crystal. Further to elucidate the magnetic ordering in Mn$_2$Sb$_2$Te$_5$, the susceptibility with its first derivative is plotted as temperature function which is in well agreement with M-T curve at 1kOe below 11K.

The linear fitting of $1/\chi$ vs T, the AFM transition corresponds to an AFM ordering which originated due to exchange coupling between $Mn^{2+}$ ions with its different nonuple layer of unit cell. The exchange coupling elucidates a Neel temperature ($T_N$) at 20 K, which is close to other reported MTIs [71,72]. The temperature dependent magnetic susceptibility of as grown $Mn_2Sb_2Te_5$ single crystal in ZFC and FC protocols at 1kOe displays an antiferromagnetic transition at $T_N \sim 20$ K. Further to elucidate the magnetic ordering, the first derivative of temperature dependent susceptibility is plotted and the $T_N$ is found to be 20K with an additional magnetic anomaly at $T_f \sim 11$ K. This is the signature of Spin-glass behavior which has short-range magnetic ordering reported in magnetic materials [73-75]. The extreme points of first derivative are carried out for the exact transition temperatures for ZFC which is shown in Fig.8(b).

In inset of fig.8(b) the reciprocal magnetic susceptibility above $T_N$ indicates the linear dependence of temperature, which follows the Curie-Weiss (CW) law

$$\frac{1}{\chi} = \frac{T - \theta_{CW}}{C}$$

where $\chi$ is the magnetic susceptibility, C is Curie constant, and $\theta_{CW}$ is the Curie Weiss temperature. The linear fitting of $1/\chi$ versus T provides the value of $\theta_{CW}$ close to -5.84 ($\theta_{CW} < 0$), elucidating ferromagnetic (FM) interactions between the $Mn^{2+}$ ions, consistent with the predicted A-type antiferromagnetic (AFM) ordering of $Mn_2Sb_2Te_5$. Similar ordering is reported by scientist for other materials $EuMg_2Sb_2$ [76], $EuSn_2As_2$[77], $SmMg_2Bi_2$[78]. Taking the relationship between curie constant and effective magnetic moment $\mu_{eff}$ defined as $C = N_A \mu_{eff}^2 / 3K_B$ which gives $\mu_{eff} = 2.828\sqrt{C}$ in unit of Bohr magneton ($\mu_B$). The spin-only $\mu_{eff}$ for high-spin $Mn^{2+}$ ions having $3d^5$ configuration is given by

$$\mu_{eff} = \sqrt{J(J+1)} * \mu_B = 5.92\mu_B \text{ where } J = 5/2.$$

In Comparison with published results by others, the previously reported magnetic moment is $4.04\mu_B$ per Mn from neutron diffraction [79], and similar to $\mu_{eff} = 5.91\mu_B$ from magnetic measurements [71]. The consistency between measured and calculated $\mu_{eff}$ is due to the intercalation of (MnTe)$_2$ magnetic layers in unit cell of Sb$_2$Te$_3$. Due to magnetic phase transition the spin fluctuates which cause deviation from the Curie-Weiss law near T$_N$ exists.

The transverse M-H curves in H∥ab plane as revealed in fig.9. elucidates apparent hysteresis loop at 5K featured by a remanent magnetization of about 1.92μ$_B$/Mn and a coercive field of about 0.006 T at 5 K. The magnetization at 20K and 100K implies a linear dependence and has no hint for saturation for T >T$_N$, indicating the paramagnetic state. The M-H curves increase slowly with field and indicate a sharp increment in magnetic moment at critical field μ$_0$H$_c$~ 0.1 T, and turn to be linear for μ$_0$H >μ$_0$H$_c$. Such behavior of M-H curve implies a phase transition into A-type AFM state. It is in good agreement with 1/χ vs T curves. In conclusion with magnetic properties, the absence of hysteresis loop in M-H curves at higher temperature is consistent with the AFM state.

**Conclusion**

In summary, Mn$_2$Sb$_2$Te$_5$ single crystals grown by self-flux method have been studied, including their structure, electrical transport properties and magnetization measurements. The SCXRD ensure that there are two major layers of MnTe in Sb$_2$Te$_3$ TIs. The unit cell of Mn$_2$Sb$_2$Te$_5$ consists nine atomic layers with c parameter to be 17.15Å. Five vibrational peaks of various frequencies are observed in Raman spectroscopy. Above 20K the linear resistivity suggesting metallic nature. The temperature dependent longitudinal resistivity and the transverse magnetization elucidating the antiferromagnetic behavior of Mn$_2$Sb$_2$Te$_5$ below 20 K with some embedded FM/PM domains at 5K and purely PM like at 100K. The magnetic moment at 1kOe is calculated to $4.8\mu_B$ which is in good agreement. Magnetic Transport feature may be useful for future research on AxI properties. As a magnetic Weyl semimetal candidate, the anomalous magnetic and transport events occurring in this study will contribute to an in-depth investigation of the topological aspects of this compound.

**Acknowledgement:**

The authors would like to thank Director NPL for his keen interest and encouragement. Ankush Saxena would like to thank Dr. Pallavi Kushwaha and Jaidev Tanwar for magnetization measurements on MPMS. Ankush Saxena would like to thank for Kamal Pandey for XPS measurements. Ankush Saxena would like to thank DST for research fellowship. Ankush Saxena would like to thanks M.M. Sharma, Yogesh Kumar for their help in laboratory work and characterization analysis. Ankush Saxena also thankful to AcSIR for Ph.D. registration.

**Table 1.**

XPS peak positions of as grown MTIs Mn$_2$Sb$_2$Te$_5$ and Full-width at half maximum of constituent elements.

| Element | Doublets | Valence States | Binding Energy (eV) | FWHM (eV) |
|---|---|---|---|---|
| Mn | 2p$_{3/2}$ | Mn$^{2+}$ | 640.94±0.02 | 1.08±0.08 |
| | | Mn$^{3+}$ | 642.28±0.04 | 2.19±0.16 |
| | 2p$_{1/2}$ | Mn$^{2+}$ | 652.86±0.10 | 1.76±0.29 |
| | | Mn$^{3+}$ | 654.33±0.11 | 2.09±0.34 |
| Sb | 3d$_{5/2}$ | Sb$^{3+}$ | 531.61±0.02 | 2.88±0.07 |
| | 3d$_{3/2}$ | Sb$^{3+}$ | 539.94±0.02 | 1.79±0.10 |
| Te | 3d$_{7/2}$ | Te$^{2-}$ | 583.59±0.01 | 1.42±0.02 |
| | 3d$_{5/2}$ | Te$^{2-}$ | 573.21±0.01 | 1.40±0.02 |

**Figure captions:**

Fig.1. Schematic of heat treatment followed to synthesize Mn$_2$Sb$_2$Te$_5$ single crystal and inset is showing the image of synthesized Mn$_2$Sb$_2$Te$_5$ single crystal.

Fig.2(a). SCXRD pattern taken on mechanically cleaved crystal flake of synthesized Mn$_2$Sb$_2$Te$_5$ single crystal.

Fig.2(b). Unit cell of synthesized Mn$_2$Sb$_2$Te$_5$ single crystal by using VESTA software.

Fig.3(a). FESEM image of synthesized Mn$_2$Sb$_2$Te$_5$ single crystal

Fig.3(b). EDX spectra of synthesized Mn$_2$Sb$_2$Te$_5$ single crystal in which the inset is showing the Stoichiometric ratio of constituent elements of Mn$_2$Sb$_2$Te$_5$.

Fig.4. De-convoluted Raman spectrum of Mn$_2$Sb$_2$Te$_5$ at room temperature.

Fig.5. XPS peaks of synthesized single crystal of Mn$_2$Sb$_2$Te$_5$ in (a) Mn 2p region (b) Sb 3d region (c) Te 3d region.

Fig.6. Illustrate normalized resistivity against temperature from 300K down to 2K at zero magnetic field.

Fig.7. MR% vs applied field of plot of ±12T synthesized $MnSb_2Te_4$ single crystal at (a) 2.5K (b) 5K, 10K, 20K, 50K, 100K and 200K.

Fig.8(a). Magnetic moment as function of temperature under the magnetic field of 100Oe and 1kOe.

Fig.8(b). First derivative of susceptibility at 1kOe and in inset the curie Weiss fitting at 1kOe.

Fig.9. Magnetic moment as function of field at 5K, 20K and 100K.

Fig.1.

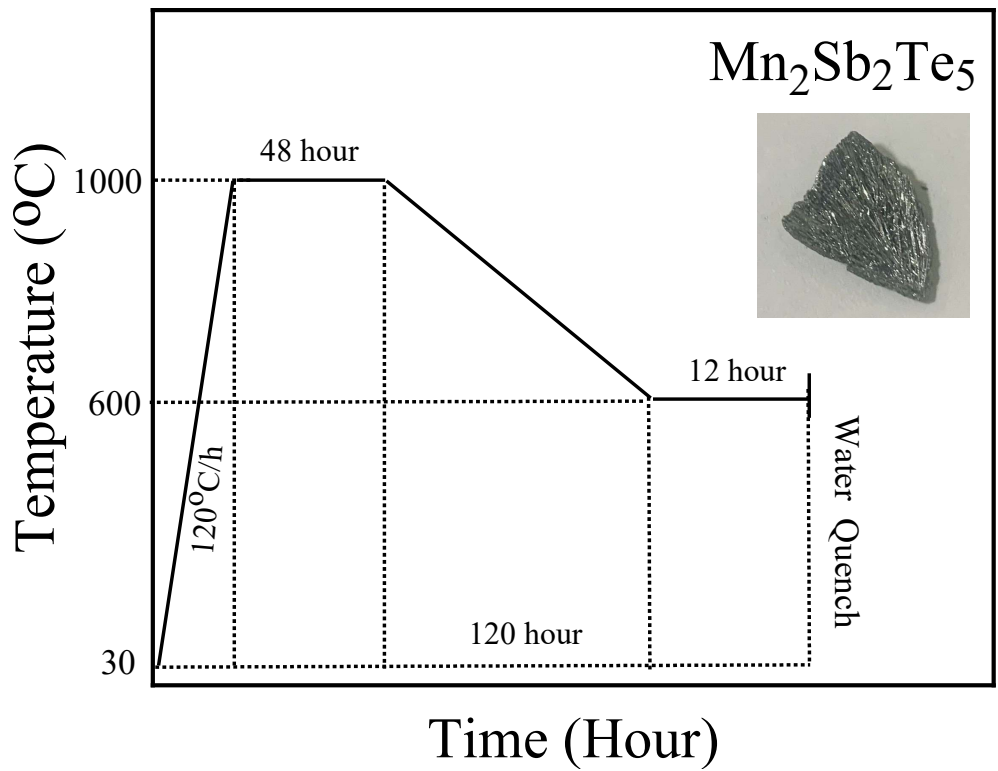

Fig. 2(a).

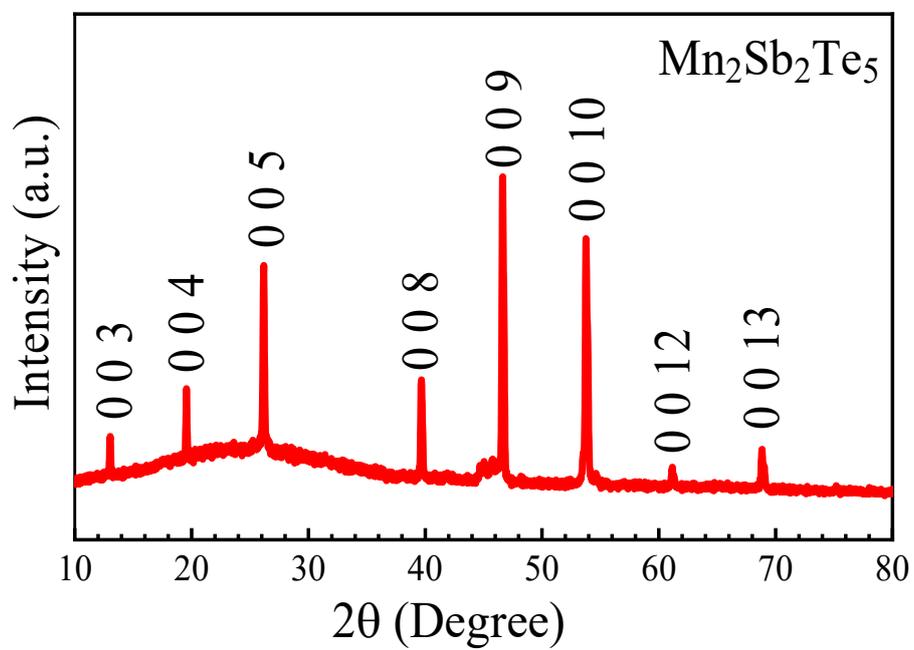

Fig. 2(b).

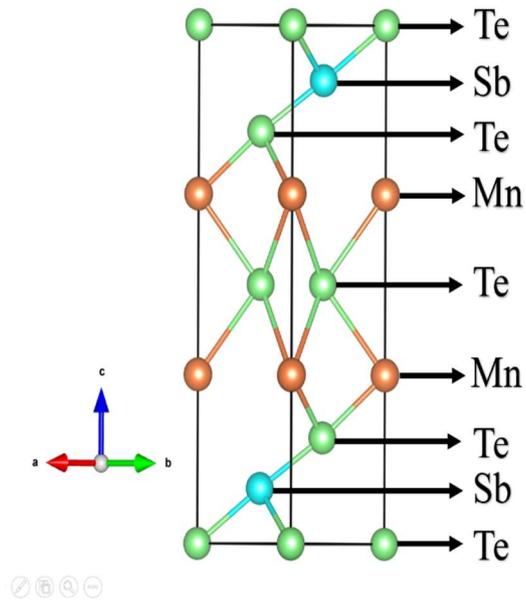

Fig. 3(a).

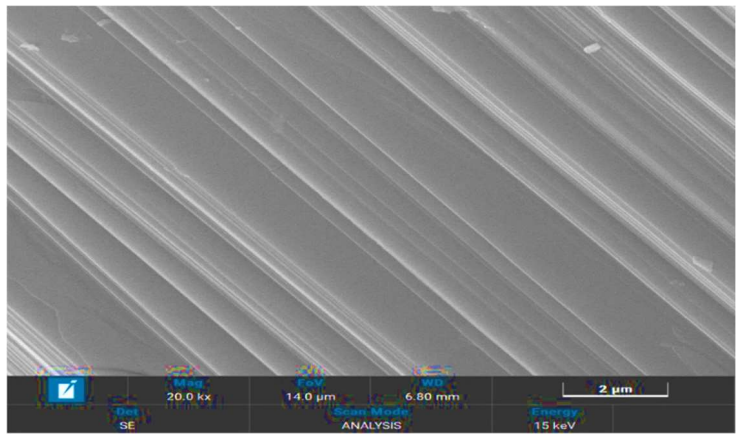

Fig. 3(b).

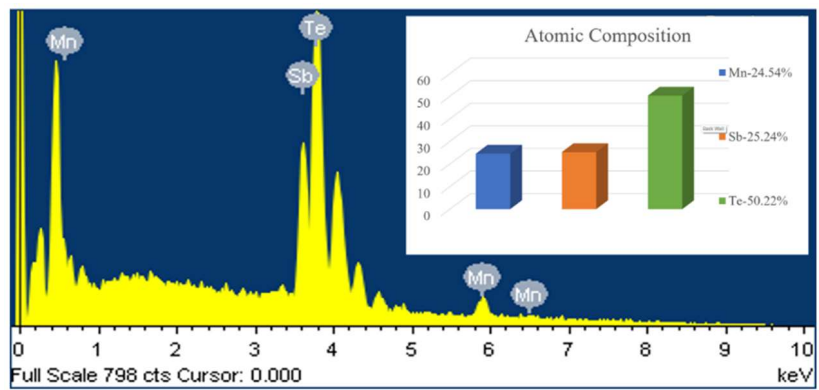

Fig.4.

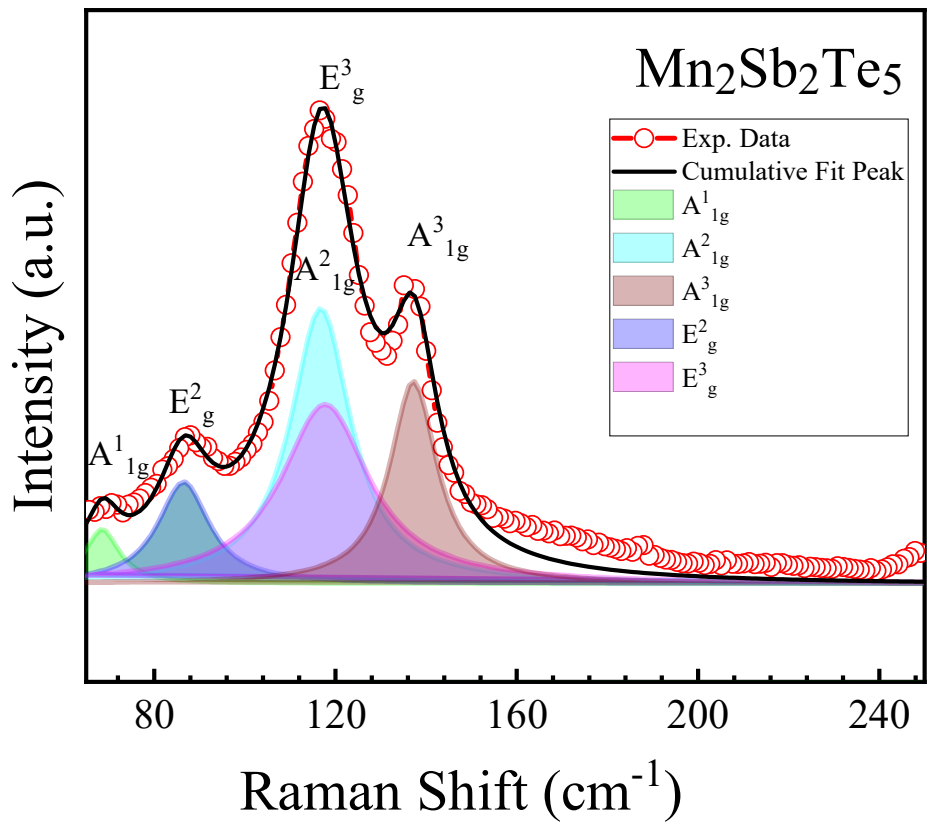

Fig.5(a).

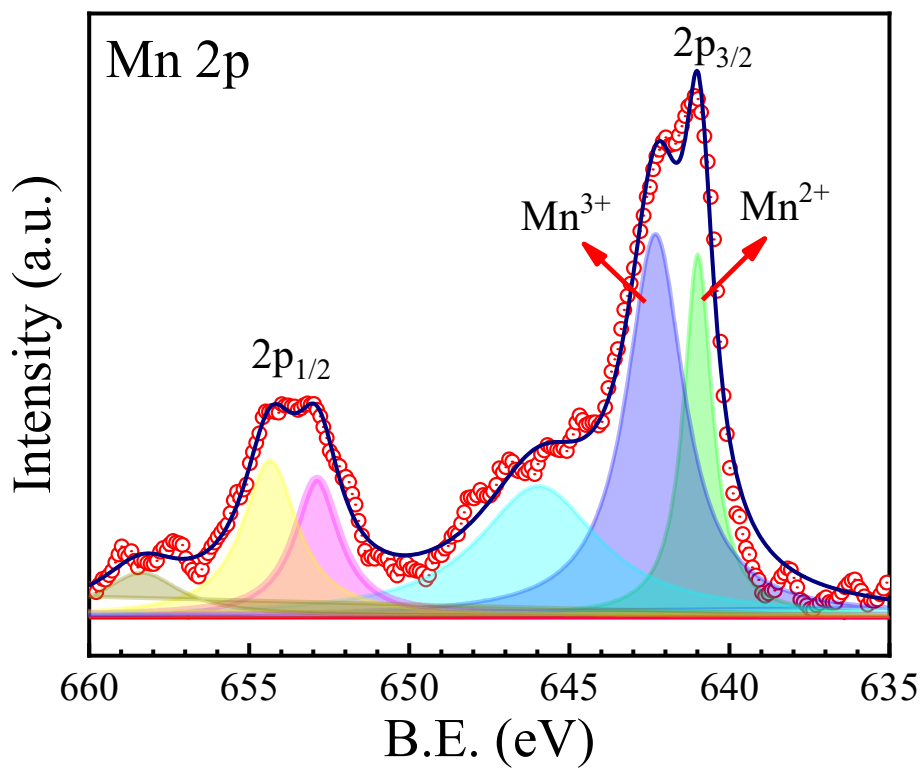

Fig.5(b).

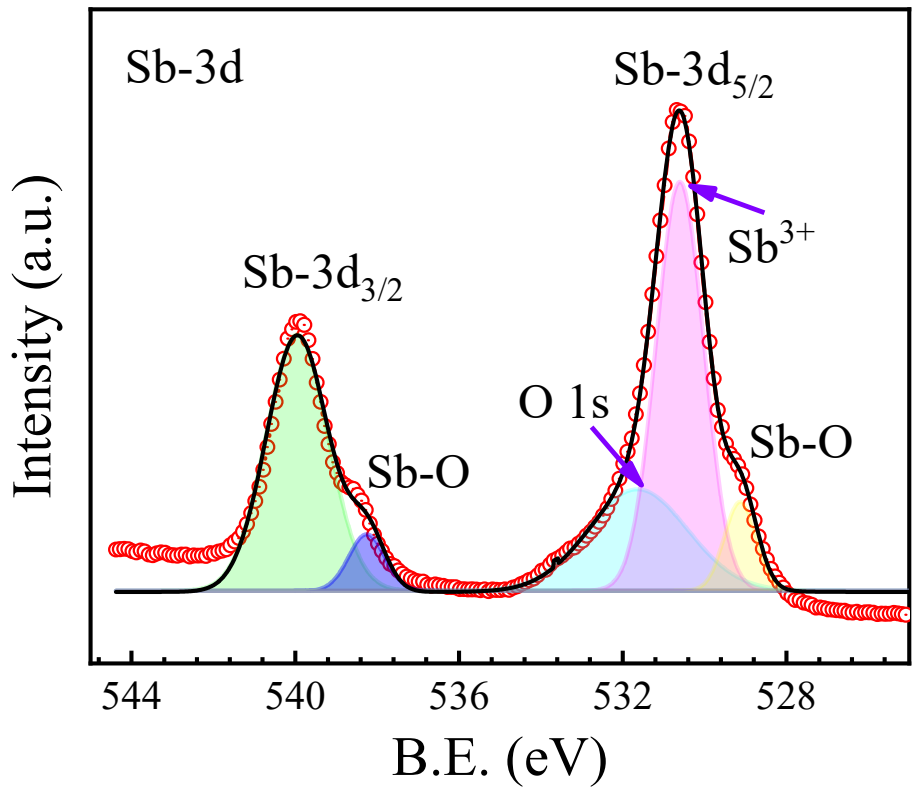

Fig.5(c).

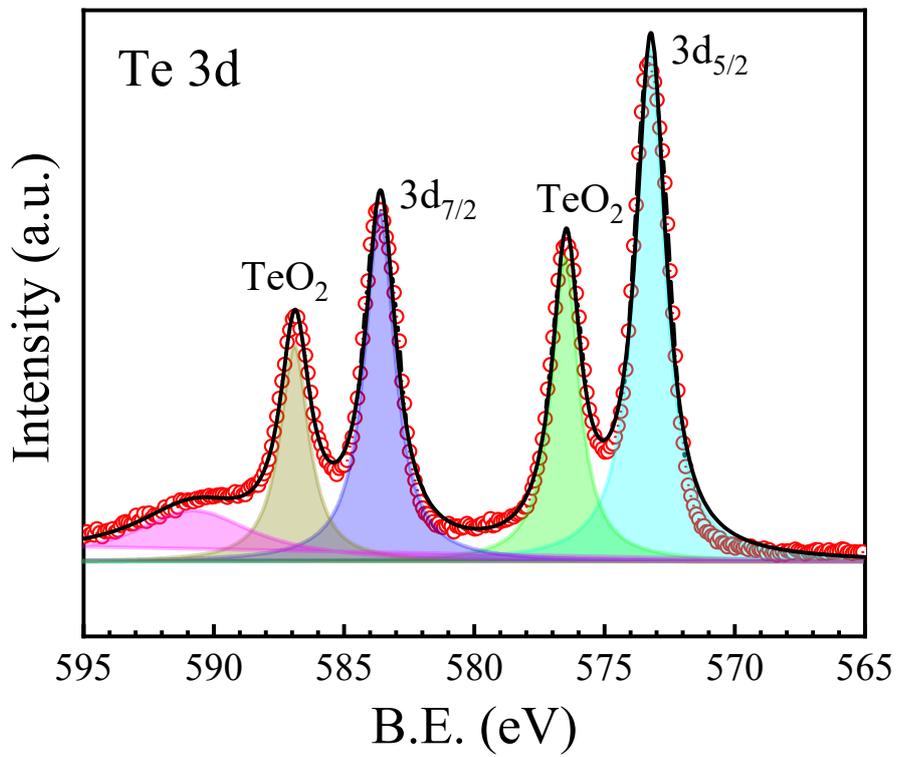

Fig.6.

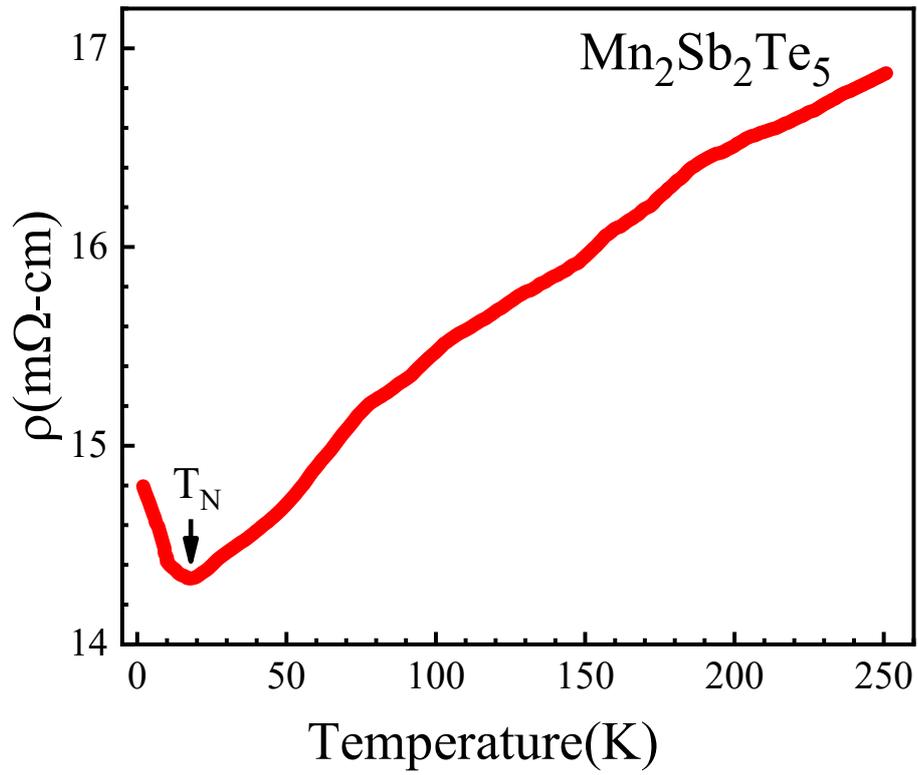

Fig.7(a).

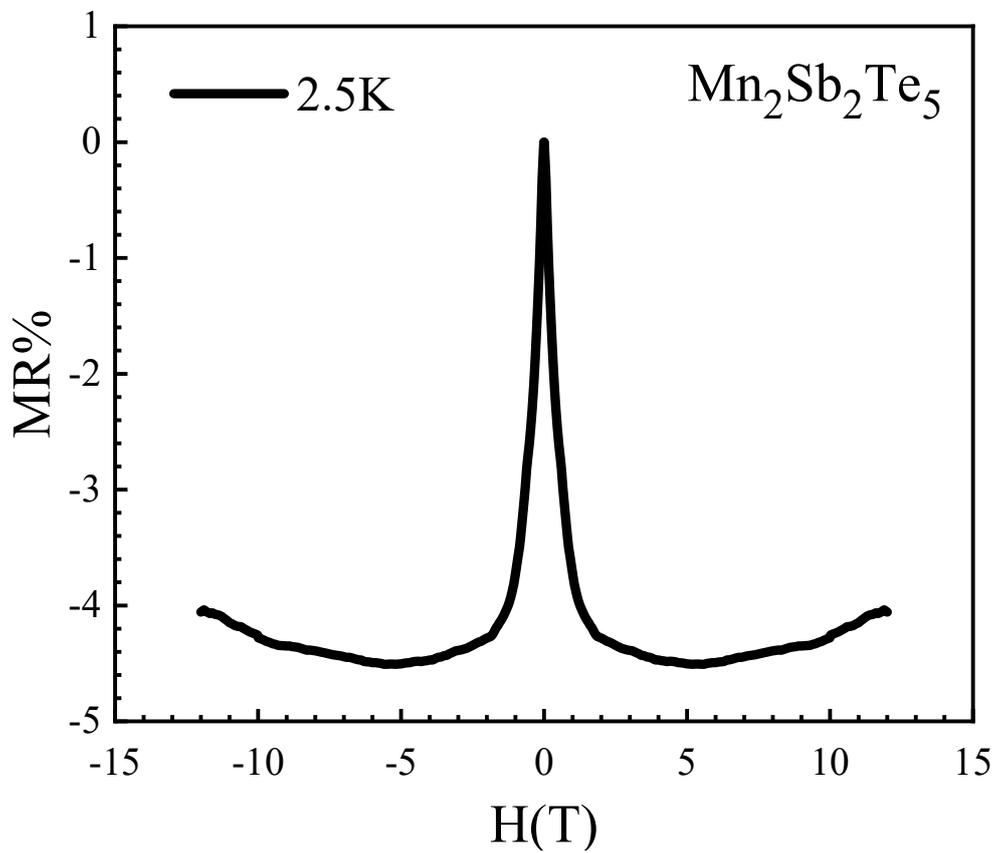

Fig.7(b).

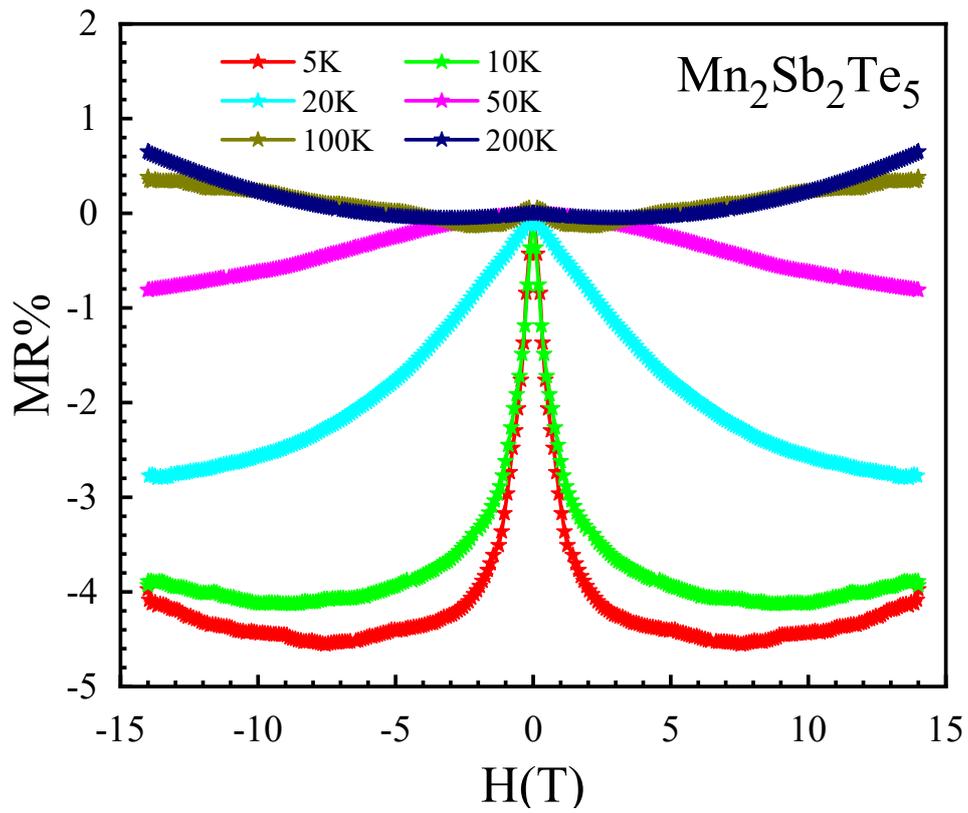

Fig.8(a).

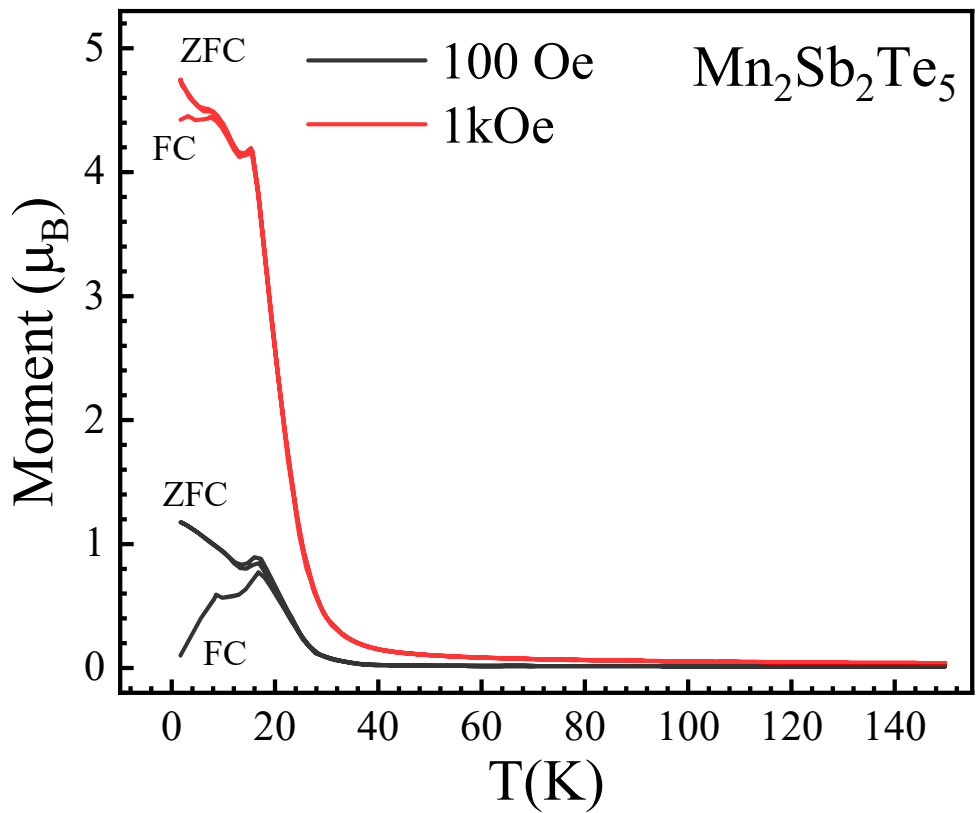

Fig.8(b).

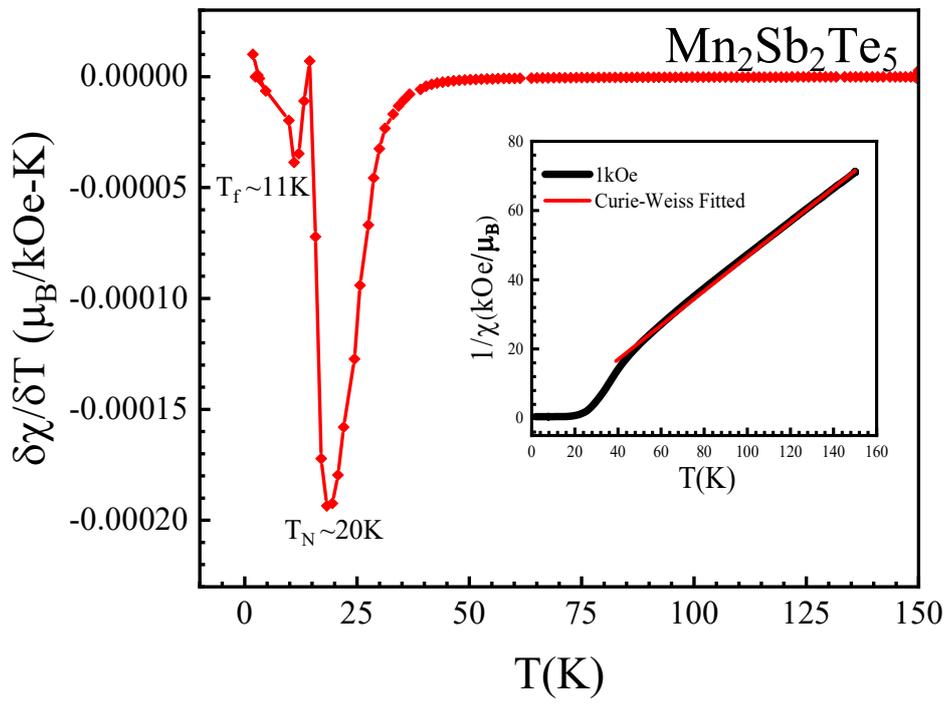

Fig.9

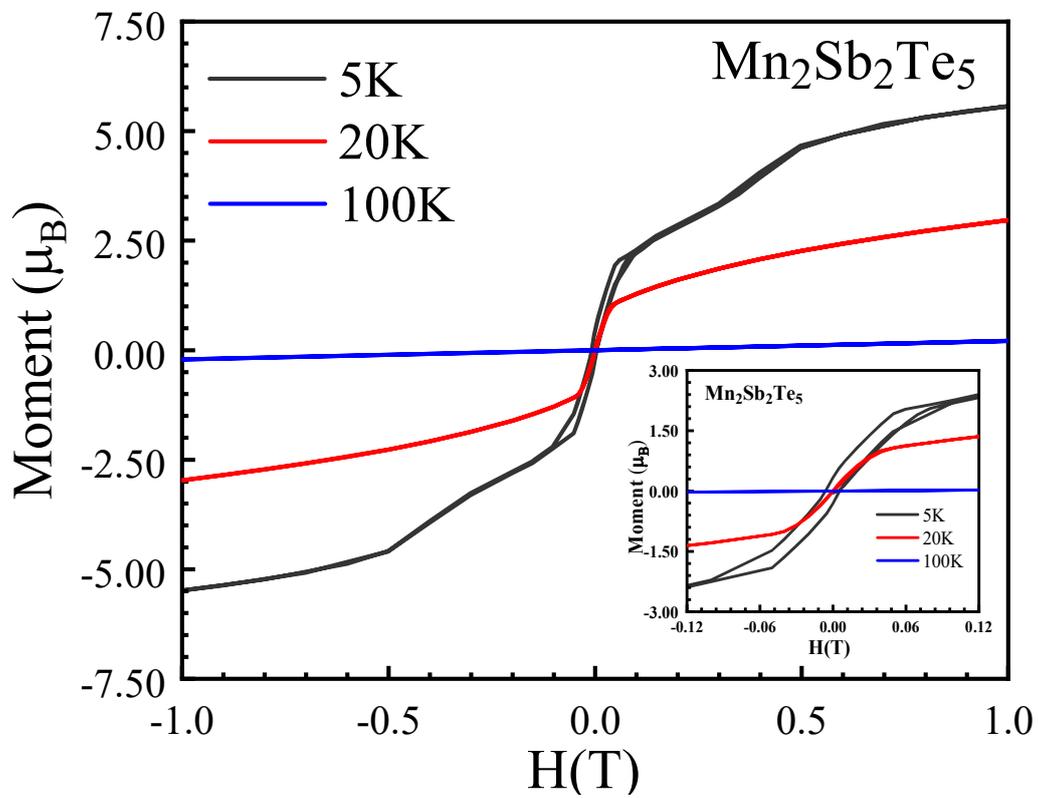

**References:**

1. F. D. M. Haldane, Phys. Rev. Lett., **61**, 2015 (1988).
2. C.-X. Liu, X.-L. Qi, X. Dai, Z. Fang, and S.-C. Zhang, Phys. Rev. Lett., **101**, 146802 (2008).
3. R. Yu, W. Zhang, H.-J. Zhang, S.-C. Zhang, X. Dai, and Z. Fang, Science., **329**, 61 (2010).
4. He, Ke. npj Quantum Mater., **5**, 90 (2020).
5. Poonam Rani, Ankush Saxena, Rabia Sultana, Vipin Nagpal, S.S. Islam, S. Patnaik, V.P.S. Awana, J. Supercond. Nov. Magnetism., **32**, 3705 (2019).
6. Chen, L., Wang, D., Shi, C. et al. J Mater Sci., **55**, 14292 (2020).
7. Gang Shi et al 2020 Chinese Phys. Lett., **37**, 047301 (2020).
8. Ankush Saxena, Poonam Rani, Vipin Nagpal, S. Patnaik, I. Felner & V. P. S. Awana, J. Sup. Nov. Mag., **33**, 2251 (2020).
9. Lin Cao, Shuang Han, Yang-Yang Lv, Dinghui Wang, Ye-Cheng Luo, Yan-Yan Zhang, Shu-Hua Yao, Phys. Rev. B., **104**, 054421 (2021).
10. Jian Zhou, Y. B. Chen, Haijun Zhang, and Yan-Feng Chen, Y. S. Hou and R. Q. Wu, Nano Lett., **19**, 2472 (2019).
11. X. Y. Wu, D. Xiao, C. Z. Chen, J. Sun, L. Zhang, M. H. W. Chan, N. Samarth, X. C. Xie, X. Lin, and C. Z. Chang, Nat. Commun., **11**, 4532 (2020).
12. R. S. K. Mong, A. M. Essin, and J. E. Moore, Phys. Rev. B., **81**, 245209 (2010).
13. C. Liu, Y. Wang, H. Li, Y. Wu, Y. Li, J. Li, K. He, Y. Xu, J. Zhang, and Y. Wang, Nat. Mater., **19**, 522 (2020).
14. D. Zhang, M. Shi, T. Zhu, D. Xing, H. Zhang, and J. Wang, Phys. Rev. Lett., **122**, 206401 (2019).
15. R. Fei, W. Song, and L. Yang, Phys. Rev. B., **102**, 035440 (2020).
16. AM. Essin and VGurarie, Phys. Rev. B., **85**, 195116 (2012).
17. B. A. Kuropatwa and H. Kleinke, Z.Anorg. Allg. Chem. **638**, 2640 (2012).
18. B. J. Kooi and J. T. M. De Hosson, J. Appl. Phys. **92**, 3584 (2002).
19. Y. Li, Y. Jiang, J. Zhang, Z. Liu, Z. Yang, and J. Wang, Phys. Rev. B **102**, 121107 (2020).
20. S. V. Eremeev, M. M. Otrokov, A. Ernst, and E. V. Chulkov,. Phys. Rev. B **105**, 195105 (2022)

21. J. L. Zhang, D. H. Wang, M. J. Shi, T. S. Zhu, H. J. Zhang, and J. Wang, Chin. Phys. Lett., **37**, 077304 (2020).
22. Rabia Sultana, Ganesh Gurjar, S Patnaik and V P S Awana, Mater. Res. Express **5**, 046107 (2018).
23. Lei Ding, Chaowei Hu, Feng Ye, Erxi Feng, Ni Ni, and Huibo Cao., Phys. Rev. B **101**, 020412 (2020)
24. N. A. Abdullaev, I. R. Amiraslanov, Z. S. Aliev, Z. A. Jahangirli, I. Yu. Sklyadneva, E. G. Alizade, Y. N. Aliyeva, M. M. Otrokov, V. N. Zverev, N. T. Mamedov & E. V. Chulkov., Jetp Lett. **115**, 749 (2022)
25. Yujin Cho, Jin Ho Kang, Liangbo Liang, Madeline Taylor, Xiangru Kong, Subhajit Ghosh, Fariborz Kargar, Chaowei Hu, Alexander A. Balandin, Alexander A. Puretzky, Ni Ni, and Chee Wei Wong, Phys. Rev. Research., **4**, 013108 (2022).
26. Wang, Xiangke Zhao, Guixia Huang, Xiubing Connor, Paul Li, Jiaxing Zhang, Shouwei Irvine, John. J. Mater. Chem. A., **3**, 297 (2014).
27. R. J. Iwanowski, M. H. Heinonen, and E. Janik, Appl. Surf. Sci., **249**, 222 (2005).
28. R. J. Iwanowski, M. H. Heinonen, and B. Witkowska, J. Alloys Compd., **491**, 13 (2010).
29. L. Zheng, X. H. Cheng, D. Cao, Q. Wang, Z. J. Wang, C. Xia, L. Y. Shen, Y. H. Yu, and D. S. Shen, RSC Adv., **5**, 40007 (2015).
30. J. Schaumann, M. Loor, D. Unal, A. Mudring, S. Heimann, U. Hagemann, S. Schulz, F. Maculewicz and G. Schierning, Dalton Trans., **46**, 656 (2017).
31. Irene Martini, Mauro Taborelli, Christoph Hessler, Eric Chevallay, Holger Neupert, V Nistor, Valentin Fedosseev., Surface Characterization at Cern of Photocathodes For Photoinjector Applications (2015).
32. Hao Li, Yaoxin Li, Yukun Lian, Weiwei Xie, Ling Chen, Jinsong Zhang, Yang Wu and Shoushan Fan, Sci. China Mater. **65**, 477 (2022).
33. Julian Schaumann, Manuel Loor, Derya Ünal, Anja Mudring, Stefan Heimann, Ulrich Hagemann, Stephan Schulz, Franziska Maculewicze and Gabi Schierning., Dalton Trans, **46**, 656 (2017)
34. Georg Bendt, Kevin Kaiser, Alla Heckel, Felix Rieger, Dennis Oing, Axel Lorke, Nicolas Perez Rodriguez, Gabi Schierning, Christian Jooss and Stephan Schulz., Semicond. Sci. Technol. **33,** 105002 (2018)
35. Hao Li, Shengsheng Liu, Chang Liu, Jinsong Zhang, Yong Xu, Rong Yu, Yang Wu, Yuegang Zhang and Shoushan Fan Phys. Chem. Chem. Phys., **22**, 556 (2020)

36. O. Gomis, R. Vilaplana, F. J. Manjon, P. Rodriguez-Hernandez, E. Perez-Gonzalez, A. Munoz, V. Kucek, and C. Drasar, Phys. Rev. B., **84**, 174305 (2011).

37. H. J. Kim, K. S. Kim, J. F. Wang, M. Sasaki, N. Satoh, A. Ohnishi, M. Kitaura, M. Yang, and L. Li, Phys. Rev. Lett. **111**, 246603 (2013).

38. K.-S. Kim, H.-J. Kim, and M. Sasaki, Phys. Rev. B **89**, 195137 (2014).

39. Q. Li, D. E. Kharzeev, C. Zhang, Y. Huang, I. Pletikosic, A. V. Fedorov, R. D. Zhong, J. A. Schneeloch, G. D. Gu, and T. Valla, Nat. Phys. **12**, 550 (2016).

40. C. Zhang et al., Nat. Commun. **7**, 10735 (2016).

41. X. C. Huang et al., Phys. Rev. X **5**, 031023 (2015).

42. J. Xiong, S. K. Kushwaha, T. Liang, J. W. Krizan, M. Hirschberger, W. Wang, R. J. Cava, and N. P. Ong, Science **350**, 413 (2015).

43. C. Z. Li, L. X. Wang, H. W. Liu, J. Wang, Z. M. Liao, and D. P. Yu, Nat. Commun. **6**, 10137 (2015).

44. C. Zhang et al., Nat. Commun. **8**, 13741 (2017).

45. H. Li, H. T. He, H. Z. Lu, H. C. Zhang, H. C. Liu, R. Ma, Z. Y. Fan, S. Q. Shen, and J. N. Wang, Nat. Commun. **7**, 10301 (2016).

46. F. Arnold et al., Nat. Commun. **7**, 11615 (2016).

47. X. J. Yang, Y. P. Liu, Z. Wang, Y. Zheng, and Z. A. Xu, arXiv:1506.03190.

48. X. Yang, Y. Li, Z. Wang, Y. Zhen, and Z.-A. Xu, arXiv:1506.02283.

49. S. L. Adler, Phys. Rev. **177**, 2426 (1969).

50. J. S. Bell and R. Jackiw, Il Nuovo Cimento A **60**, 47 (1969).

51. H. B. Nielsen and M. Ninomiya, Nucl. Phys. B **185**, 20 (1981).

52. J. Wang et al., Nano Res. **5**, 739 (2012).

53. H. T. He, H. C. Liu, B. K. Li, X. Guo, Z. J. Xu, M. H. Xie, and J. N. Wang, Appl. Phys. Lett. **103**, 031606 (2013).

54. S. Wiedmann et al., Phys. Rev. B **94**, 081302 (2016).

55. L.-X. Wang, Y. Yan, L. Zhang, Z.-M. Liao, H.-C. Wu, and D.-P. Yu, Nanoscale **7**, 16687 (2015).

56. O. Breunig, Z. Wang, A. A. Taskin, J. Lux, A. Rosch, and Y. Ando, Nat. Commun. **8**, 15545 (2017).

57. B. A. Assaf, T. Phuphachong, E. Kampert, V. V. Volobuev, G. Bauer, G. Springholz, L. A. de Vaulchier, and Y. Guldner, Phys. Rev. Lett. **119**, 106602 (2017).

58. Bergmann G. Phys. Rep. **107** 1(1984)

59. Patrick A. Lee and T. V. Ramakrishnan., Rev. Mod. Phys. **57** 287(1985)

60. Ankush Saxena, N.K. Karn, M.M. Sharma, V.P.S. Awana, J. Phys. Chem. of Solids., **174**, 111169 (2023).

61. Ankush Saxena, M.M. Sharma, Prince Sharma, Yogesh Kumar, Poonam Rani, M. Singh, S. Patnaik, V.P.S. Awana, J. Alloys Compd., **895**, 162553 (2022).

62. Gang Shi, Mingjie Zhang, Dayu Yan, Honglei Feng, Yang Meng, Youguo Shi, Yongqing Li., Anomalous Hall Effect in Layered Ferrimagnet $MnSb_2Te_4$, (2020).

63. Peng-Fei Zhu, Xing-Guo Ye, Jing-Zhi Fang, Peng-Zhan Xiang, Rong-Rong Li, Dai-Yao Xu, Zhongming Wei, Jia-Wei Mei, Song Liu, Da-Peng Yu, and Zhi-Min Liao., Phys. Rev. B **101**, 075425 (2020)

64. N. E. Sluchanko, A. L. Khoroshilov, M. A. Anisimov, A. N. Azarevich, A. V. Bogach, V. V. Glushkov, S. V. Demishev, V. N. Krasnorussky, N. A. Samarin, N. Yu. Shitsevalova, V. B. Filippov, A. V. Levchenko, G. Pristas, S. Gabani, and K. Flachbart., Phys. Rev. B **91**, 235104 (2015)

65. N. J. Harmon and M. E. Flatte, Phys. Rev. Lett. **108**, 186602 (2012)

66. Binder K, Young AP. Rev Mod Phys, **58** 801 (1986)

67. Mydosh JA. Spin Glasses: An Experimental Introduction. London: Taylor & Francis, (1993).

68. Bin Pang, Lunyong Zhang, Y. B. Chen, Jian Zhou, Shuhua Yao, Shantao Zhang, and Yanfeng Chen ACS Applied Materials & Interfaces, **9**, 3201 (2017)

69. Adhip Agarwala and Vijay B. Shenoy. Phys. Rev. Lett. **118**, 236402 (2017)

70. F. Bern, M. Ziese, A. Setzer, E. Pippel, D. Hesse, I. Vrejoiu. J. Phys. Condens. Matter, **25**, 496003 (2013)

71. J.Q. Yan, Q. Zhang, T. Heitmann, Z. L. Huang, W. D. Wu, D. Vaknin, B. C. Sales, and R. J. McQueeney. Phys. Rev. Mater. 3, 064202 (2019).

72. Liu, C. Wang, Y. Li H. Wu Y. Li Y. Li J. He, K. Xu, Y. Zhang J. Wang Y., arXiv:1905.00715 (2019).

73. T. H. and Y. O. D X Li, A Dönni, Y Kimura, Y Shiokawa, Y Homma, Y Haga, E Yamamoto, J. Phys. Condens. Matter **11**, 8263 (1999).

74. A.V. Semeno, M.A. Anisimov, A.V. Bogach., S.V. Demishev, M. I. Gilmanov,V. B. Filipov, N.Yu. Shitsevalova, V.V.Glushkov., Sci Rep **10**, 18214 (2020).

75. S.H. Kim, P.D. Battle., J. Mag. Mag. Mater. **123**, 273 (1993)

76. Santanu Pakhira, Farhan Islam, Evan O'Leary, M. A. Tanatar, Thomas Heitmann, Lin-Lin Wang, R. Prozorov, Adam Kaminski, David Vaknin, and D. C. Johnston., Phys. Rev. B **106**, 024418 (2022)

77. Santanu Pakhira, M. A. Tanatar, Thomas Heitmann, David Vaknin, and D. C. Johnston Phys. Rev. B **104**, 174427 (2021)
78. Asish K. Kundu, Santanu Pakhira, Tufan Roy, T. Yilmaz, Masahito Tsujikawa, Masafumi Shirai, E. Vescovo, D. C. Johnston, Abhay N. Pasupathy, and Tonica Valla Phys. Rev. B **106**, 245131 (2022)
79. Zeugner, A Nietschke, F. Wolter, A. U. B. Gass, S. Vidal, R. C. Peixoto, T. R. F. Pohl, D. Damm, C. Lubk, A. Hentrich, R. Moser, S. K. Fornari, C. Min, C. H. Schatz, S. Kissner, K. Unzelmann, M. Kaiser, M. Scaravaggi, F. Rellinghaus, B. Nielsch, K. Hess, C. Buchner, B. Reinert, F. Bentmann, H. Oeckler, O. Doert, T. Ruck, M. Isaeva, A., Chem. Mater. **31**, 8 (2019).